\documentclass[aps,prl,twocolumn,showpacs]{revtex4}
\begin{document}
\title{Quantum Entanglement in Second-quantized Condensed Matter Systems}
\author{Yu Shi}

\affiliation{Department of Physics, University of Illinois at
Urbana-Champaign, Urbana, IL 61801, USA}

J. Phys. A {\bf 37}, 6807 (2004)

\begin{abstract}

The entanglement between occupation-numbers of different single
particle basis states depends on coupling between different single
particle basis states in the second-quantized Hamiltonian. Thus in
principle, interaction is not necessary for occupation-number
entanglement to appear. However, in order to characterize {\em
quantum correlation caused by interaction}, we use the eigenstates
of the single-particle Hamiltonian as the single particle basis
upon which the occupation-number entanglement is defined. Using
this so-called proper single particle basis, if there is no
interaction, then the many-particle second-quantized Hamiltonian
is diagonalized and thus cannot generate entanglement, while its
eigenstates can always be chosen to be non-entangled. If there is
interaction, entanglement in the proper single particle basis
arises in energy eigenstates and can be dynamically generated.
Using the proper single particle basis, we discuss
occupation-number entanglement in  important eigenstates,
especially  ground states, of systems of many identical particles,
in exploring insights the notion of entanglement sheds on
many-particle physics. The discussions on Fermi systems start with
Fermi gas, Hatree-Fock approximation, and the electron-hole
entanglement in excitations. In the ground state of a Fermi
liquid, in terms of the Landau quasiparticles, entanglement
becomes negligible. The entanglement in a quantum Hall state is
quantified as $-f\ln f-(1-f)\ln (1-f)$, where $f$ is the proper
fractional part of the filling factor. For BCS superconductivity,
the entanglement is a function of the relative momentum
wavefunction of the Cooper pair $g_{\mathbf{k}}$, and is thus
directly related to the superconducting energy gap, and vanishes
if and only if superconductivity vanishes. For a spinless Bose
system, entanglement does not appear in the
Hatree-Gross-Pitaevskii approximation, but becomes important in
the Bogoliubov theory, as a characterization of two-particle
correlation caused by the weak interaction. In these examples, the
interaction-induced entanglement as calculated is directly related
to the macroscopic physical properties.
\end{abstract}


\maketitle

\section*{1. Introduction}

Quantum entanglement is the situation that a quantum state of a
composite system is not a direct product of states of the
subsystems~\cite{einstein}.  It is an essential quantum feature
without classical analogy~\cite{bell,mermin}. For many decades,
the notion of entanglement has been mostly used in foundations of
quantum mechanics. Recently it was found to be crucial in quantum
information processing. For a bipartite pure state
$|\psi_{AB}\rangle$, the entanglement can be quantified as the von
Neumann entropy of the reduced density matrix of either party,
$S=-tr_A\rho_A\ln\rho_A=-tr_B\rho_B\ln \rho_B$, where
$\rho_A=tr_B(|\psi_{AB}\rangle\langle\psi_{AB}|)$,
$\rho_B=tr_A(|\psi_{AB}\rangle\langle\psi_{AB}|)$~\cite{bennett}.
Thus  $0\leq S \leq \ln D$, where $D$ is  the smaller one of the
dimensions of the Hilbert spaces of $A$ and $B$. The larger $S$,
the stronger the entanglement. Recall that the von Neumann entropy
of a density matrix is a measure of the distribution of  its
eigenvalues; the more homogeneous this distribution, the larger
the von Neumann entropy.

Since quantum entanglement is an essential quantum correlation, it
is natural and interesting to consider useful or even fundamental
insights that the notion of entanglement may provide on quantum
many-body physics and quantum field theory.  Historically, similar
consideration was made in Yang's study of off-diagonal long-range
order~\cite{yang} and in Leggett's study of
disconnectivity~\cite{leg}.  The recent development of quantum
information theory may be useful to some important issues in
frontiers of physics~\cite{preskill}. Some investigations have
been made on entanglement between  spins at different sites in
some spin lattice
models~\cite{hei1,thermal,osterloh,osborn,vidal,korepin1,korepin2}.
Nevertheless, the bulk of quantum many-body physics concerns
identical particles, with the localized spin models as special
cases. Hence in this regard, it is inevitable to address the issue
of entanglement in systems of identical particles. This topic,
related to both quantum information and condensed matter physics,
is pursued in various approaches
\cite{abrams,lloyd1,lloyd2,bravyi,mn,terhal,zanardi1,zanardi0,shi4,shi0,dis,enk,vedral1,ghirardi,knill,schliemann}.

As the quantum correlation beyond the symmetrization or
anti-symmetrization of identical particles, entanglement between
occupation numbers of different single particle basis states
(modes) is an appropriate characterization.  Lloyd and coworkers
when recognized the occupation number basis as the suitable basis
for quantum simulation of second quantized many-particle systems,
exemplified by using the Hubbard model~\cite{abrams}. Furthermore,
occupation was also proposed as the degree of freedom to implement
qubit~\cite{lloyd1,lloyd2,bravyi,mn,terhal}. Zanardi noted the
isomorphism between the full Fock space and qubits space and
investigated the entanglement in grand canonical
ensembles~\cite{zanardi1}. Afterwards, from a physical standpoint
and the relation between occupation number state and the
(anti)symmetrized particle state, the present author carefully
justified the use of Fock space in investigating entanglement
issue, even in the case of particle number conservation, thus
helped to establish the applicability of this approach to
many-particle pure states~\cite{shi4,shi0}. It is also noted that
the magnetic spin entanglement is a special case of
occupation-number entanglement of identical particles~\cite{shi4}.
Some related papers have appeared after the present work was
actually done~\cite{shi0}. Vedral made some very interesting
investigations, where the application of two-mode squeezing to
Bose condensates and related systems were analyzed, and
two-particle fermionic entanglement due to symmetrization were
computed~\cite{vedral1}.

In this approach, clearly the entanglement in the many-particle
system depends on which single particle basis is chosen. This
point might seem uncomfortable to some researchers, since they
remember that entanglement should not be affected by local
operations. Let us emphasize that the subsystems are defined by
modes, not by particles. The choice of single particle basis
defines how to partition the system into subsystems, and actually
defines the single particles~\cite{zeh}.  {\em Once the single
particle basis is chosen, i.e. the partition into subsystems is
defined, the entanglement is invariant under unitary operations on
individual subsystems, i.e. the modes, in fully consistent with
the general wisdom about entanglement}. Naturally a question
arises: which single particle basis does one choose? The answer is
that it depends on the relevance to the question one is concerned,
or which single particle basis corresponds to the particles that
are detected in the circumstance.

Given that in this approach, the entanglement is between the
occupation-numbers of different single particle basis states,
whether it can be generated by, or whether it exists in an
eigenstate of, a many-particle second-quantized Hamiltonian ${\cal
H}$ depends on whether there is coupling between different single
particle basis states in ${\cal H}$, in contrast with the case of
distinguishable particles, for which the entanglement depends on
interaction of particles. Hence generically speaking, interaction
is {\em not} necessary for generation or existence of
occupation-number entanglement. For example, in a single particle
basis in which ${\cal H}$  is not diagonal, the occupation-number
entanglement exists in eigenstates of  ${\cal H}$, and can be
dynamically generated.

Here we note, however, there is a special single particle basis in
which, if there is no interaction, entanglement cannot be
generated from a non-entangled state, while each {\em energy
eigenstate} must be non-entangled (except the insignificant case
of a superposition of two degenerate non-entangled eigenstates
with different occupation numbers in at least two single particle
basis states). In this single particle basis, entanglement in a
non-degenerate energy eigenstate is only caused by interaction.
Hence this single particle basis is very suitable in
characterizing the {\em quantum correlation due to interaction},
rather than the entanglement that appears merely as a consequence
of Bogoliubov mode transformation.   For convenience, let us call
this special single particle basis  {\em the proper single
particle basis}.

The so-called proper single particle basis is just the set of
eigenstates of the single particle Hamiltonian. It is indexed by
the (continuous or discrete) momentum in the case of free
particles, the Bloch wave-vector plus the band index in the case
of particles in a periodic potential, the degree of the Hermit
polynomial and the perpendicular momentum in the case of electrons
in a magnetic field, etc. The inclusion of spin as an additional
index is straightforward.

It is instructive and amusing to consider our method of
characterizing interaction-induced entanglement as an extension of
the novel way of counting states of a system of identical
particles invented by Bose, Einstein and Dirac~\cite{pais}. They
considered ideal gas, hence the underlying many-particle states
are just all the possible occupation-number basis states in the
momentum basis, which is the proper single particle basis in this
question.  Each of these occupation-number basis states is a
direct product of the occupation states of single particle basis
states (modes). No superposition of these occupation-number basis
states. No entanglement between the proper single particle basis
states. Hence the classical Boltzmann counting is applicable when
one considers the occupations of the single particle states,
rather than the particles themselves. The entanglement between
these different single particle states emerges when there is
interaction, as discussed in this paper.

In this paper, using the proper single particle basis, we shall
explore the {\em interaction-induced}  entanglement in
representative many-particle states, which are of fundamental
importance in condensed matter physics and the like. In
particular, we emphasize the role of Hamiltonian and the relation
between entanglement with macroscopic physical properties.

Energy eigenstates, especially the ground states, are of utmost
importance in many-body and statistical physics. Besides,
adiabatically controlled ground states is also used in  some
quantum computing schemes~\cite{farhi,zanardi2,sw}. Hence it is
important to address the issue of entanglement in the energy
eigenstates, especially the ground state. These  aspects further
motivate our work.

The organization of this paper is the following. First an
introduction and clarification is made on occupation-number
entanglement in a system of many identical particles, and
especially to the so-called proper single particle basis. Then we
discuss the ground state and excitations of normal Fermi systems,
especially the electron-hole entanglement in the Hatree-Fock
approach. In the next two sections, we make detailed
investigations on entanglement in quantum Hall effect and
Bardeen-Cooper-Schrieffer (BCS) superconductivity, respectively.
Afterwards there is a section on Bosonic entanglement, in which
entanglement in Bogoliubov theory is calculated. We summarize
after making some additional remarks.

\section*{2. The proper single particle basis}

In the standard formalism of  second quantization,  one can write
a  state of many identical particles in terms of an arbitrarily
chosen single particle basis,  as
\begin{equation}
|\psi\rangle = \sum_{n_1,\cdots, n_{\infty}}
 f(n_1,\cdots,n_{\infty})|n_1,\cdots,n_{\infty}\rangle, \label{pn}
\end{equation}
where  $n_i$ is the occupation number of single particle state $i$
in the chosen single particle basis, $|n_i\rangle \equiv
(1/\sqrt{n_i !}) {a_i^{\dagger}}^{n_i}|0\rangle$
$|n_1,\cdots,n_{\infty}\rangle$ corresponds to a Slater
determinant or permanent wavefunction in the configuration space.
For a fixed number of particles, whether a many-particle state is
entangled means whether the wavefunction is a single Slater
determinant or permanent. In principle, entanglement in a system
of identical particles is a property dependent on which single
particles and which single particle basis is chosen in
representing the many-particle system, and can be quantified as
that among occupation numbers of different single particle states.

Choosing a different single particle basis means partitioning the
system into a different set of subsystems, based on which the
entanglement is then defined. But once a single particle basis is
chosen, the entanglement in invariant under any unitary operation
on individual single particle basis states, i.e. when there is no
coupling between different single particle basis states. In other
words, in the present case,  the meaning of ``local operations''
as previously used in quantum information theory is generalized to
operations on the corresponding single particle basis states, as
indexed by the subscript $i$ above. Of course, it is constrained
that some kinds of generalized ``local'' unitary operations do not
exist physically. Once this generalization of the meaning of
subsystems and local operations is made, the usual method of
calculating the amount of entanglement, as developed in quantum
information theory, can be applied.

Quantitatively, one considers the Fock-state reduced density
matrix of a set of single particle basis states  $1,\cdots, l$,
\begin{equation}
\begin{array}{c}
 \langle n_1',\cdots,n_l'|\rho_l(1\cdots
l)|n_1,\cdots,n_l\rangle
\equiv \nonumber \\
\sum_{n_{l+1},\cdots,n_{\infty}} \langle n_1',\cdots,n_l',
n_{l+1},n_{\infty}|\rho|n_1,\cdots,n_l, n_{l+1},n_{\infty}\rangle.
\end{array}
 \label{reduce}
\end{equation}
Its von Neumann entropy measures the entanglement of this set of
single particle basis states and the rest of the system, relative
to the empty state.  If the total number of particles is
conserved, then it is constrained that the only matrix elements
which may be nonzero are those with $\sum_{i=1}^l n_i'
=\sum_{i=1}^l n_i$. In particular, the reduced density matrix of
one single particle basis state is always diagonal, indicating
entanglement whenever there are more than one nonzero diagonal
elements.

In this approach, the statistics determines the dimensions of the
Hilbert space of each mode. For fermions, $n_i=0,1$, $D=2$, hence
the entanglement between one single particle basis state and the
rest of the system is $0\leq S\leq \ln 2$. For bosons, $n_i$ is
arbitrary, hence $D$ is infinity. This point does not pose real
difficulties, but further investigation on it is interesting.

One can also define the entanglement relative to the ground state,
by considering only the effect of creation and annihilation
operators acting on the ground state. Then $n_i$ in (\ref{reduce})
is understood as the number of  the excited particles, which are
absent in the ground state $|G\rangle$, i.e.
$|n_1,\cdots,n_{\infty}\rangle \equiv (1/\sqrt{n_1!\cdots
n_{\infty}!}) {a_1^{\dagger}}^{n_1} \cdots
{a_{\infty}^{\dagger}}^{n_{\infty}}|G\rangle$.

The non-relativistic  field theoretic or second quantized Hamiltonian
is
\begin{equation}
\begin{array}{ll}
{\cal H} & =  \int
d^3r\hat{\psi}^{\dagger}(\mathbf{r})h(\mathbf{r})
\hat{\psi}(\mathbf{r})+ \int
d^3r\hat{\psi}^{\dagger}(\mathbf{r})h'(\mathbf{r})
\hat{\psi}(\mathbf{r})\nonumber
\\ &  + \frac{1}{2}\int d^3r \int d^3r'
\hat{\psi}^{\dagger}(\mathbf{r})\hat{\psi}^{\dagger}(\mathbf{r}')
V(\mathbf{r},\mathbf{r}')\hat{\psi}(\mathbf{r}')\hat{\psi}(\mathbf{r}),
\label{field} \end{array}
\end{equation}
where $h(\mathbf{r})$ is the single particle Hamiltonian including
the kinetic energy, $V(\mathbf{r},\mathbf{r}')$ is the
particle-particle interaction, $h'(\mathbf{r})$ is some external
potential which is not included in $h(\mathbf{r})$ for
convenience.  In the examples in this paper, $h' \neq 0 $ only in
the issue of generating electron-hole excitations by
electron-light interaction; $h'=0$ in all discussions on
entanglement in many-particle energy eigenstates. The field
operator $\hat{\psi}(\mathbf{r})$ can be expanded in an
arbitrarily chosen single particle basis as
$\hat{\psi}(\mathbf{r})=\sum_i\phi_i(\mathbf{r})a_i$, where $i$ is
the collective index of the single particle state,   which may
include spin if needed, $a_i$ is the annihilation operator,
$\phi_i(\mathbf{r})$ is the single particle wavefunction in
position space. We use the same notations for fermions and bosons.
Thus ${\cal H}$ can also be written as
\begin{equation}
\begin{array}{lll}
{\cal H} & = & \sum\limits_{ij} \langle i|h|j\rangle a_i^{\dagger}a_j \\
& & + \sum\limits_{ij}
 \langle i|h'|j\rangle a_i^{\dagger}a_j + \frac{1}{2}\sum\limits_{ijlm} \langle
ij|V|lm\rangle a_i^{\dagger}a_j^{\dagger} a_ma_l, \end{array}
\label{sec}
\end{equation}
The generalization to the existence of more species of identical
particles is straightforward. Single particle basis transformation
leads to a unitary transformation in the creation and annihilation
operators.  There may  be more general transformations of the
creation and annihilation operators, and some even involve
combination of operators of different species. Such a
transformation means describing the system in terms of a different
set of single particles or quasiparticles.

Even if $V=0$ and $h'=0$, as far as $\langle i|h|j\rangle \neq 0$,
${\cal H}$ can  generate occupation-number entanglement between
single particle basis state $i$ and $j$.

An eigenstate of a second quantized interacting Hamiltonian is
often entangled.  In the chosen single particle basis, if an
eigenstate of (\ref{sec}) is non-entangled, then it must be of the
form $|\psi\rangle = \otimes_i |n_i\rangle$. Consequently for each
$i$, ${\cal H}\hat{n}_i|\psi\rangle = \hat{n}_i {\cal
H}|\psi\rangle$, where $\hat{n}_i =a_i^{\dagger}a_i$. It can be
seen that this is often not satisfied by ${\cal H}$ in
(\ref{sec}).

However,  when we use entanglement to characterize the {\em
quantum correlation caused by interaction}, it is suitable to use
the set of eigenstates of the single particle Hamiltonian $h$,
which we call proper single particle basis. In this single
particle basis, with $h\phi_\mu =\epsilon_{\mu}\phi_\mu$,  $\int
d^3r\hat{\psi}^{\dagger}(\mathbf{r})h(\mathbf{r})
\hat{\psi}(\mathbf{r})= \sum_{\mu} \epsilon_{\mu}
a_{\mu}^{\dagger}a_{\mu}$, whose eigenstates are of the form
$\otimes_{\mu}|n_{\mu}\rangle$, where $\mu$ is the collective
index of the proper single particle basis.

Therefore in the proper single particle basis,  entanglement can
be used to characterize the effect of interaction. In case $h'=0$,
it characterizes the effect of the particle-particle interaction.
Each non-degenerate energy eigenstate of the non-interacting
system must be non-entangled. When there is degeneracy, an
entangled energy eigenstate of a free system may be constructed as
a superposition of degenerate non-entangled states that differ in
the occupation-numbers of  at least two single particle basis
states (on the other hand, particle number conservation constrains
that it is impossible to be different only in one single particle
basis state). But one can always use a set of non-entangled
eigenstates. If in the proper single particle basis, entangled
energy eigenstates inevitably arise, then there must be
interaction.

Besides, the proper single particle basis directly corresponds to
the energy spectrum of single particle excitations, and is more
experimentally accessible.

For the so-called strongly correlated systems, e.g. Luttinger
liquid and fractional quantum Hall state discussed below, peculiar
physical properties are caused by the strong (Coulomb)
interaction, hence it is particularly interesting to consider
occupation-number entanglement in the proper single particle
basis. By generalizing the method to relativistic field theory, it
may be useful for quantum chromodynamics.

On the other hand, when an improper single particle basis is used,
even the one-body term in ${\cal H}$ is not diagonal, and the
eigenstates are entangled even when there is no interaction, as
seen by transforming $a_{\mu}^{\dagger}$ in
$\otimes_{\mu}|n_{\mu}\rangle\equiv\otimes_{\mu}
 (1/\sqrt{n_{\mu}!}) {a_{\mu}^{\dagger}}^{n_{\mu}}|0\rangle $. Nevertheless,
entanglement in an improper basis may be interesting in problems
such as hopping, tunnelling, Mott transition, etc. For example, in
a two-state problem, of which the double well potential problem is
an example, the proper basis states are linear superpositions of
the two states, but in many cases it is these two states that are
observed.  As occupation-number entanglement in an improper single
particle basis presents even when there is no interaction, it may
be valuable for quantum information processing.

When there are more than one index in the single particle basis,
one of them can be used as the tag effectively distinguishing the
particles, and the other indices determine whether they are
entangled in these degrees of freedom. With this effective
distinguishability, the state in the configuration space of  the
remaining degrees of freedom can be directly obtained from the
second-quantized state. For example, in $\frac{1}{\sqrt{2}}(
a_{\mathbf{k}'\uparrow}^{\dagger}a_{\mathbf{k}\downarrow}^{\dagger}
+
a_{\mathbf{k}'\downarrow}^{\dagger}a_{\mathbf{k}\uparrow}^{\dagger})
|0\rangle$, where $\mathbf{k}'$ and $\mathbf{k}$ represent
momenta, one can say that the particle in $|\mathbf{k}'\rangle$
and the particle in $|\mathbf{k}\rangle$ are spin-entangled. One
can also say that the particle in $|\uparrow\rangle$ and the
particle in $|\downarrow\rangle$ are momentum-entangled. With the
momentum as the distinguishing tag, the spin state is
$\frac{1}{\sqrt{2}}(|\uparrow\rangle_{\mathbf{k}'}|\downarrow\rangle_{\mathbf{k}}+
|\uparrow\rangle_{\mathbf{k}'}|\downarrow\rangle_{\mathbf{k}})$.
Alternatively, with the spin as the distinguishing tag, the
momentum state is
$\frac{1}{\sqrt{2}}(|\mathbf{k}'\rangle_{\uparrow}|\mathbf{k}\rangle_{\downarrow}
+|\mathbf{k}\rangle_{\uparrow}|\mathbf{k}'\rangle_{\downarrow})$.

The ideas about the occupation-number entanglement can be
consistently applied even to a one-particle state
$|\phi\rangle=\sum_i c_i|i\rangle$, where $|i\rangle$'s are a set
of basis states. In terms of occupation numbers of different basis
states, the state can be written as $|\phi\rangle = \sum_i c_i
|1\rangle_i\prod_{j\neq i}|0\rangle_j$. Thus the occupation-number
of basis state $|i\rangle$ is entangled with other basis states,
with the  amount of entanglement $-|c_i|^2\ln |c_i|^2
-(1-|c_i|^2)\ln(1-|c_i|^2)$. When $|\phi\rangle$ and $|i\rangle$
's are eigenstates of the Hamiltonian, $|\phi\rangle =|I\rangle$,
thus $c_I=1$ while $c_j=0$ for $j\neq I$, consequently in
$|\phi\rangle$,  each basis state is non-entangled with other
basis states.  In the example of an electron in a superposition of
a state $|-\mathbf{k}'\rangle_e$ in a Fermi sea and a state
$|\mathbf{k}\rangle_e$ out of the Fermi sea, written in terms of
 the occupation-numbers of these two electronic states,
$a|0\rangle_{e\mathbf{k}}|0\rangle_{e-\mathbf{k}'}+
b|1\rangle_{e\mathbf{k}}|1\rangle_{e-\mathbf{k}'}$,  can also be
written in terms of occupation-numbers of  the electron state
$|\mathbf{k}\rangle_e$ and the  hole state
$|\mathbf{k}'\rangle_h$, as
$a|0\rangle_{e\mathbf{k}}|0\rangle_{h\mathbf{k}'}+b|1\rangle_{e\mathbf{k}}
|1\rangle_{h\mathbf{k}'}$. This becomes a superposition of absence
and presence of an electron-hole pair. But this kind of
electron-hole entanglement is different from  the entanglement
between an existing electron and an existing hole.

Now we start our discussions on occupation-number entanglement in
important energy eigenstates in many-particle physics, using the
so-called proper single particle basis. These systems play
fundamental roles in condensed matter physics.

\section*{3. Fermi systems}

First let us consider a fermi gas, which plays a fundamental role
in understanding condensed matter physics. The proper single
particle basis here is the tensor product of single-particle
momentum and spin states.

The ground state of a free Fermi gas is $|G\rangle=
\prod_{\mathbf{k}}^{|{\mathbf{k}}| < k_F}
a_{\mathbf{k}\uparrow}^{\dagger}a_{\mathbf{k}\downarrow}^{\dagger}|0\rangle$,
where $k_F$ is the Fermi momentum. It is clearly non-entangled. An
excited state like
$a_{{\mathbf{k}}s}^{\dagger}b_{{\mathbf{k}}'s'}^{\dagger}|G\rangle$
is still separable, where $|{\mathbf{k}}|>k_F>|{\mathbf{k}'}|$,
$b^{\dagger}_{{\mathbf{k}}s}=a_{-{\mathbf{k}}-s}$ is the hole
operator. It is simple to check that for each of these
non-entangled states, a Fock-space reduced density matrix, as in
(\ref{reduce}), always has only one nonzero element, hence the
entanglement between the occupation-numbers of any set of single
particle basis states  and the rest of the system indeed vanishes.

There may be entanglement in an excited state of a Fermi gas,
because of degeneracy due to spin degree of freedom. For example,
there is maximal entanglement in the electron-hole pair
$\frac{1}{\sqrt{2}}(
a_{\mathbf{k}\uparrow}^{\dagger}b_{\mathbf{k}'\downarrow}^{\dagger}
+a_{\mathbf{k}\downarrow}^{\dagger}b_{\mathbf{k}'\uparrow}^{\dagger})
|G\rangle$. For a free gas, it is a superposition of  the
degenerate non-entangled states
$a_{\mathbf{k}\uparrow}^{\dagger}b_{\mathbf{k}'\downarrow}^{\dagger}
|G\rangle$ and
$a_{\mathbf{k}\downarrow}^{\dagger}b_{\mathbf{k}'\uparrow}^{\dagger}
|G\rangle$.  With respect to the empty state, it is  an
entanglement between the occupation-number of the excited electron
state and others. Since $|G\rangle$ is non-entangled, entanglement
in an excited state relative to the empty state is equal to the
entanglement relative to $|G\rangle$.  The entanglement in
$\frac{1}{\sqrt{2}}(a_{\mathbf{k}\uparrow}^{\dagger}
b_{\mathbf{k}'\downarrow}^{\dagger}
+a_{\mathbf{k}\downarrow}^{\dagger}b_{\mathbf{k}'\uparrow}^{\dagger})
|G\rangle$ is simply electron-hole entanglement with respect to
the ground state. Moreover,  an electron and a hole, by
definition, corresponds to different single particle states, and
can be regarded as distinguishable particles, as tagged by that a
creation operator of a hole  corresponds to  annihilation of an
electron.  In the absence of interaction, however, one can always
use a set of non-entangled energy eigenstates as the orthonormal
set.

More realistic treatment, in the context of solid state physics,
takes into account the Coulomb interaction between the electrons,
as well as the crystal structure, which provides a single particle
(periodic) potential. A basic method is the Hatree-Fock
approach~\cite{solid}. The ground state is still non-entangled,
since the Hatree-Fock treatment only modifies the single particle
states and ground state energy.  But entanglement inevitably
arises in excited states. To illustrate the idea, the simplest
model of electronic excitations in solids is considered in the
following.

Consider one electron is excited from a valence band to a
conduction band. An eigenstate of this excitation, an exciton, is
$\sum_{\mathbf{k},\mathbf{k}'}
A_{\mathbf{k},\mathbf{k}'}a_{\mathbf{k}}^{\dagger}
b_{\mathbf{k}'}^{\dagger}|G\rangle$, in the spinless case. For
brevity, the band indices are omitted, as one corresponds to the
electron operator while the other corresponds to the hole
operator.  The occupation numbers of  the basis states
$|\mathbf{k}\rangle_e$ and $|\mathbf{k}'\rangle_h$ respectively
occupied by the excited electron and by the hole are the same as
those in relative to the ground state, since they are zero in the
ground state. The Fock-space reduced density matrix elements of
$\mathbf{k}$ can be obtained  as $\langle
1|\rho_1(\mathbf{k})|1\rangle = \alpha_\mathbf{k}\equiv
\sum_{\mathbf{k}'}|A_{\mathbf{k},\mathbf{k}'}|^2$, $\langle
0|\rho_1(\mathbf{k})|0\rangle = 1-\alpha_\mathbf{k}$. Therefore
the occupation-number entanglement between the electron basis
state $|\mathbf{k}\rangle_e$  and the rest of the system is
$-\alpha_\mathbf{k}\ln\alpha_{\mathbf{k}}-
(1-\alpha_\mathbf{k})\ln(1-\alpha_{\mathbf{k}})$. The
occupation-number entanglement between the hole basis state
$|\mathbf{k}'\rangle_h$ and the rest of the system is
$-\alpha_{\mathbf{k}'}\ln\alpha_{\mathbf{k}'}-
(1-\alpha_{\mathbf{k}'})\ln(1-\alpha_{\mathbf{k}'})$,  where
$\alpha_{\mathbf{k}'} =
\sum_{\mathbf{k}}|A_{\mathbf{k},\mathbf{k}'}|^2$. The Fock-space
reduced density matrix $\rho_{1,1}$ of the electron basis state
$|\mathbf{k}\rangle_e$  plus the hole basis state
$|\mathbf{k}'\rangle_h$ as a subsystem is calculated by
considering that the electron and the hole belong to different
species of identical particles. $\langle
1,1|\rho_{1,1}(\mathbf{k},\mathbf{k}')|1,1\rangle =
|A_{\mathbf{k},\mathbf{q}'}|^2$, $\langle
1,0|\rho_{1,1}(\mathbf{k},\mathbf{k}')|1,0\rangle =
\gamma_{\mathbf{k}}\equiv \sum_{\mathbf{q}'\neq
\mathbf{k}'}|A_{\mathbf{k},\mathbf{q}'}|^2$, $\langle
0,1|\rho_{1,1}(\mathbf{k},\mathbf{k}')|0,1\rangle =
\gamma_{\mathbf{k}'}\equiv \sum_{\mathbf{q}\neq
\mathbf{k}}|A_{\mathbf{q},\mathbf{k}'}|^2$. Furthermore,
$\rho_{1,1}(\mathbf{k},\mathbf{k}')$ must be diagonal. Hence their
occupation-number entanglement with the rest of the system is
$-|A_{\mathbf{k},\mathbf{k}'}|^2 \ln|A_{\mathbf{k},\mathbf{k}'}|^2
-\gamma_{\mathbf{k}}\ln\gamma_{\mathbf{k}}
-\gamma_{\mathbf{k}'}\ln\gamma_{\mathbf{k}'}
-(1-|A_{\mathbf{k},\mathbf{k}'}|^2
-\gamma_{\mathbf{k}}-\gamma_{\mathbf{k}'}) \ln(1-
|A_{\mathbf{k},\mathbf{k}'}|^2-
\gamma_{\mathbf{k}}-\gamma_{\mathbf{k}'})$.

With the electron and the hole effectively distinguishable, the
state can be written, in the configuration space, as
$\sum_{\mathbf{k},\mathbf{k}'}
A_{\mathbf{k},\mathbf{k}'}|\mathbf{k}\rangle_e|\mathbf{k}'\rangle_h$.
The entanglement between these two distinguishable particles is
obtained by finding the eigenvalues of the reduced density matrix
for either particle.

With spin degeneracy, the  excitonic states  are
$\sum_{\mathbf{k},\mathbf{k}'}
A_{\mathbf{k},\mathbf{k}'}|S,S_z\rangle_{\mathbf{k},\mathbf{k}'}$,
where  $|S,S_z\rangle_{\mathbf{k},\mathbf{k}'}$ represents three
triplet states as the ground states,
$|1,1\rangle_{\mathbf{k},\mathbf{k}'} =
a_{\mathbf{k}\uparrow}^{\dagger}b_{\mathbf{k}'\uparrow}^{\dagger}|G\rangle$,
$|1,0\rangle_{\mathbf{k}\mathbf{k}'} =
\frac{1}{\sqrt{2}}(a_{\mathbf{k}\uparrow}^{\dagger}b_{\mathbf{k}'\downarrow}^{\dagger}
-a_{\mathbf{k}\downarrow}^{\dagger}b_{\mathbf{k}'\uparrow}^{\dagger})|G\rangle$
and $|1,-1\rangle_{\mathbf{k},\mathbf{k}'} =
a_{\mathbf{k}\downarrow}^{\dagger}
b_{\mathbf{k}'\downarrow}^{\dagger}|G\rangle$,
and one singlet state $|0,0\rangle_{\mathbf{k},\mathbf{k}'} =
\frac{1}{\sqrt{2}}(a_{\mathbf{k}\uparrow}^{\dagger}b_{\mathbf{k}'\downarrow}^{\dagger}
+a_{\mathbf{k}\downarrow}^{\dagger}
b_{\mathbf{k}'\uparrow}^{\dagger})|G\rangle$.

The occupation-number entanglement, with the full collective index
including Bloch wavevector and spin, can be calculated in a way
similar to the spinless case. The above discussions on the
spinless case applies similarly to $\sum_{\mathbf{k},\mathbf{k}'}
A_{\mathbf{k},\mathbf{k}'}|1,\pm
1\rangle_{\mathbf{k},\mathbf{k}'}$.  For
$\sum_{\mathbf{k},\mathbf{k}'}
A_{\mathbf{k},\mathbf{k}'}\frac{1}{\sqrt{2}}
(a_{\mathbf{k}\uparrow}^{\dagger}b_{\mathbf{k}'\downarrow}^{\dagger}
\pm
a_{\mathbf{k}\downarrow}^{\dagger}b_{\mathbf{k}'\uparrow}^{\dagger})|G\rangle$,
one can find, for example, that  the occupation-number
entanglement between the electron basis state
$|\mathbf{k},\uparrow\rangle_e$  plus the hole basis state
$|\mathbf{k}',\downarrow\rangle_h$ as a subsystem and the rest of
the system is $- (|A_{\mathbf{k},\mathbf{k}'}|^2/2) \ln
(|A_{\mathbf{k},\mathbf{k}'}|^2/2)
-(\gamma_{\mathbf{k}}/2)\ln(\gamma_{\mathbf{k}}/2)
-(\gamma_{\mathbf{k}'}/2)\ln(\gamma_{\mathbf{k}'}/2)-
(1-\gamma_{\mathbf{k}}/2-\gamma_{\mathbf{k}'}/2-
|A_{\mathbf{k},\mathbf{k}'}|^2/2)
\ln(1-\gamma_{\mathbf{k}}/2-\gamma_{\mathbf{k}'}/2-
|A_{\mathbf{k},\mathbf{k}'}|^2/2)$, and that the occupation-number
entanglement between the electron basis state
$|\mathbf{k},\uparrow\rangle_e$  plus the hole basis state
$|\mathbf{k}',\uparrow\rangle_h$ as a subsystem and the rest of
the system is $-(\alpha_{\mathbf{k}}/2)\ln(\alpha_{\mathbf{k}}/2)
-(\alpha_{\mathbf{k}'}/2)\ln(\alpha_{\mathbf{k}'}/2)
-(1-\alpha_{\mathbf{k}}/2-\alpha_{\mathbf{k}'}/2)
\ln(1-\alpha_{\mathbf{k}}/2-\alpha_{\mathbf{k}'}/2)$.

The entanglement considered above is determined by the Coulomb
interaction, as $A_{\mathbf{k},\mathbf{k}'}$ is determined by the
Schr\"{o}dinger equation in momentum representation. When Coulomb
interaction is negligible, $A_{\mathbf{k},\mathbf{k}'}=1$ for a
specific pair of values of $\mathbf{k}$ and $\mathbf{k}'$, and
consequently various entanglements concerning the $\mathbf{k}$ and
$\mathbf{k}'$ as discussed above consistently vanish. The spin
part of the eigenstates can be chosen to be non-entangled.
Interaction causes spread of $A_{\mathbf{k},\mathbf{k}'}$ and thus
non-vanishing entanglement in the Bloch wavevectors, as well as
the spin-entanglement. Noteworthy is that the detail of the
interaction only affects $A_{\mathbf{k},\mathbf{k}'}$, but does
not affect the structure of the spin states.

With the electron and the hole effectively distinguishable, the
state can be written, in the configuration space, as
$\sum_{\mathbf{k},\mathbf{k}'}
A_{\mathbf{k},\mathbf{k}'}|\mathbf{k}\rangle_e|\mathbf{k}'\rangle_h|S,S_z\rangle$.
So the orbital and spin degrees of freedom are actually separated,
as consistent with the fact that the spin-orbit coupling has been
neglected here.

An excited state is often generated by electron-light interaction
switched on for a period.  The light is treated as classical. The
electron-light interaction corresponds to $h'$ in Section 2. With
coupling between different single electron and hole basis states,
it can generate electron-hole entanglement. This underlies  a
recent experimental  result~\cite{chen}, on which a theoretical
analysis, with spin-orbit coupling taken into account, will be
given elsewhere~\cite{sl}.

If an interacting fermi system can be described as a Fermi liquid,
then there is a one-to-one correspondence between the particles in
the non-interacting system and the quasi-particles of the
interacting  system, obtained by adiabatically turning on the
interaction~\cite{landau}. Therefore in terms of the
quasiparticles, the ground state of a Fermi liquid is
non-entangled. One may say that the electron entanglement caused
by the interaction can be renormalized away. In contrast, the
ground state of Luttinger liquid is a global unitary
transformation of a Fermi sea~\cite{luttinger}, and is entangled.
New ground states emerge in phenomena like quantum Hall effect and
superconductivity, in which entanglement is important, as shown in
the next two sections.

\section*{4. Quantum Hall Effect}

Quantum Hall states are obtained by filling  the spin polarized
electrons in the degenerate (single particle) Landau
levels~\cite{qhe}. The single particle Hamiltonian, corresponding
to the proper single particle basis, is the Hamiltonian of a
two-dimensional electron in a magnetic field. One knows that the
degeneracy of each Landau level, i.e. the number of different
states of each energy eigenstate, is the same. The key quantity in
quantum Hall effect is the filling factor $\nu$, which is the
number of electrons divided by the degeneracy of each Landau
level, and manifested in the quantized Hall resistivity.

We show that the filling factor $\nu$  determines the
entanglement. First, the entanglement vanishes in integer quantum
Hall effect, which appears when $\nu=n$ is an integer. The $n$
lowest Landau levels are completely filled while others are empty.
Because of energy gap, the interaction is not important, and the
ground state of the interacting systems can be smoothly connected
to that of the non-interacting system. Thus the ground state is
just a product state $\prod_{\mu}a_{\mu}^{\dagger}|0\rangle$,
where $\mu$ runs over the filled states. Hence the occupation
number of each single particle state belonging to a completely
filled Landau levels is $1$, while the occupation number of every
other single particle state is $0$. In the Fock-space reduced
density matrix of a single particle basis state, for a state $\mu$
belonging to a completely filled Landau level, only $\langle 1
|\rho_1(\mu)|1\rangle=1$ is nonzero, while for a state $\mu$
belonging to an empty Landau level, only $\langle 0 |\rho_1(\mu)
|0\rangle=1$ is nonzero. It is like the ground state of a free
Fermi gas. All single particle basis states are separable from one
another, and there is no entanglement.

In a fractional quantum Hall state of $\nu=n+f$, where $f$ is the
proper fractional part,  $n\geq 0$ lowest Landau levels are
completely filled, $f$ of the next Landau level is filled, the
higher Landau level are empty. Because of partial filling, the
interaction cannot be treated perturbatively, and in the ground
state, electrons are strongly correlated.  Each single particle
basis state belonging to one of the $n$ completely filled levels
or the empty levels is separated, i.e. the occupation number is
either $0$ or $1$ and is just a factor in the many-particle state
in the particle number representation.

For each single particle basis state in the partially filled
landau level, its entanglement with the rest of the system is
obtained as follows. Consider the identity
\begin{equation}
\langle 1|\rho_1(\mu)|1\rangle = \sum_{ n_1 \cdots n_{\infty} } n_{\mu}
\langle n_1\cdots n_{\infty}|\rho|n_1\cdots n_{\infty}\rangle
=\langle \hat{n}_{\mu}\rangle,
\end{equation}
where $n_{\mu}=0,1$ ($\mu =1,\cdots,\infty$), $\langle \hat{n}_{\mu}\rangle
\equiv  Tr(\rho\hat{n}_{\mu})$ is the expectation value of the
particle number at state $\mu$. The first equality is valid only for
fermions while the second is valid for both fermions and boson. On
the other hand, in an isotropic  uniform state, for each single
particle basis state belonging to the partially filled Landau
level,
\begin{equation}
\langle \hat{n}_{\mu} \rangle  = f.
\end{equation}
Therefore the entanglement between a single particle basis state
belonging to the partially filled Landau level and the rest of the
system is
\begin{equation}
S= -f\ln f-(1-f)\ln(1-f).
\end{equation}

This simple expression of entanglement, in terms of the proper
fractional part of filling factor, gives a useful measure of the
quantum correlation in a quantum Hall state. The entanglement
increases from $0$ at $f=0$, corresponding to the integer quantum
Hall effect, towards the maximum $\ln 2$, after which it decreases
towards $0$ at $f=1$, corresponding to the integer quantum Hall
effect again. Note that the filling factor, hence the
entanglement, is extremely precisely measured, with topological
stability. Anyons have been proposed as a candidate to implement
fault-tolerant quantum computing~\cite{lloyd2,kitaev}. The present
result confirms the intrinsic entanglement, which is needed for
quantum computing.

The amount of entanglement obtained above  is consistent with the
fact that for an integer quantum Hall effect, the many-particle
wavefunction is  a single Slater determinant, indicating
separability,  while for fractional quantum effect, it is
not~\cite{laughlin}, indicating the existence of entanglement. The
Laughlin state is indeed isotropic uniform.

The fractional quantum Hall effect can be understood in terms of
the composite fermions or composite bosons~\cite{qhe}, which are
non-entangled. For example,  the state at $\nu=1/(2p+1)$ can be
viewed as $\nu_{eff}=1$ integer quantum Hall state of composite
fermions, or equivalently as Bose condensation of composite
bosons, while $\nu=1/2p$ state is a free Fermi gas with a Fermi
surface. In each of these descriptions, the system of the
composite particles is separable. This separability can also be
inferred from the off-diagonal long-range order (ODLRO) exhibited
by these composite particles~\cite{girvin}, because
disentanglement of the condensate mode from other modes underlies
ODLRO~\cite{dis}. ODLRO is an important notion in many-particle
physics, and is the hallmark of Bose condensation and
superconductivity~\cite{yang}.

\section*{5. BCS Superconductivity}

As another example of using the concept of  entanglement to
further our understanding of many-particle  physics, we now
consider BCS superconductivity~\cite{bcs,schrieffer}. The
Hamiltonian is ${\cal H}=\sum_{\mathbf{k},s}
\epsilon_{\mathbf{k}}n_{\mathbf{k},s}+\sum_{\mathbf{k},\mathbf{k}'}
\langle \mathbf{k}',-\mathbf{k}'|V|\mathbf{k},-\mathbf{k}\rangle
a_{\mathbf{k}'\uparrow}^{\dagger}a_{-\mathbf{k}'\downarrow}^{\dagger}
a_{-\mathbf{k}\downarrow}a_{\mathbf{k}\uparrow}$. The proper
single particle basis is $(\mathbf{k},s)$, in which the one-body
term in ${\cal H}$ is diagonalized. The BCS superconducting ground
state is,
\begin{equation}
|\psi_0\rangle = {\cal N}_0  \prod_{\mathbf{k}} (1+g_{\mathbf{k}}
a_{\mathbf{k}\uparrow}^{\dagger}a_{-\mathbf{k}\downarrow}^{\dagger})
|0\rangle,
\end{equation}
where ${\cal N}_0= \prod_ {\mathbf{k}}
(1+|g_{\mathbf{k}}|^2)^{-1/2}$ is the normalization factor. For
$|\psi_0\rangle$, in which the particle number is not conserved,
 The entanglement is only between each pair
$({\mathbf{k}},s)$ and $(-{\mathbf{k}},-s)$, with the amount
\begin{equation}
S_0= -z_{\mathbf{k}}\ln z_{\mathbf{k}}-(1-z_{\mathbf{k}})\ln
(1-z_{\mathbf{k}}),
\end{equation}
where $z_{\mathbf{k}} =1/(1+|g_{\mathbf{k}}|^2)$. There is no
entanglement between different pairs.

However, for a system of $N$ electrons, with $N$ fixed,  the exact
state is the projection of $|\psi_0\rangle$ onto the $N$-particle
space, which is~\cite{schrieffer}\footnote{Without loss of
essence,  suppose $N$ to be even here. It is straightforward to
extend to the  case in which the number of particles is odd or the
system is  at an excited state.  For example, if  one particle at
$(\mathbf{p},\uparrow)$ has no mate $(-\mathbf{p},\downarrow)$,
$|\psi_0\rangle$ is replaced as $ \prod_{\mathbf{k}\neq
\mathbf{p}} (1+g_{\mathbf{k}}a_{\mathbf{k}\uparrow}^{\dagger}
a_{-\mathbf{k}\downarrow}^{\dagger})
a_{\mathbf{p},\uparrow}|0\rangle$}.
\begin{equation}
|\psi(N)\rangle
 = {\cal N}_N
\sum g_{{\mathbf{k}}_{1}}
a_{{\mathbf{k}}_1\uparrow}^{\dagger}a_{-{\mathbf{k}}_1\downarrow}^{\dagger}
\cdots g_{{\mathbf{k}}_{N/2}}
a_{{\mathbf{k}}_{N/2}\uparrow}^{\dagger}
a_{-{\mathbf{k}}_{N/2}\downarrow}^{\dagger}
|0\rangle,
 \label{nstate}
\end{equation}
where  ${\cal N}_N=(\sum  |g_{{\mathbf{k}}_1}|^2\cdots
|g_{{\mathbf{k}}_{N/2}}|^2)^{-1/2}$, $\sum$  represents summations
over ${\mathbf{k}}_1, \cdots, {\mathbf{k}}_{N/2}$, with the
constraint ${\mathbf{k}}_i\neq {\mathbf{k}}_{j}$. One can observe
that this state  is given by the superposition of all kinds of
products of  $N/2$ different $g_{\mathbf{k}}
a_{\mathbf{k}\uparrow}^{\dagger}a_{-\mathbf{k}\downarrow}^{\dagger}$.
This feature  leads to entanglement between different  Cooper
paired modes.

Let us investigate the entanglement in $|\psi(N)\rangle$. First we
evaluate the elements of the Fock-space
 reduced density matrix of mode $({\mathbf{k}},s)$, denoted
 as  $\langle n_{{\mathbf{k}},s}|\rho_1({\mathbf{k}},s)
|n_{{\mathbf{k}},s}\rangle$.  One can obtain
$$\langle 1_{{\mathbf{k}},s}|\rho_1({\mathbf{k}},s)
|1_{{\mathbf{k}},s}\rangle = x_{\mathbf{k}}=
\frac{|g_{\mathbf{k}}|^2 {\sum}' |g_{\mathbf{k}_2'}|^2\cdots
|g_{\mathbf{k}_{N/2}'}|^2}{\sum
 |g_{{\mathbf{k}}_1}|^2\cdots
|g_{{\mathbf{k}}_{N/2}}|^2},$$
 where $\sum'$ represents  the summations
over ${\mathbf{k}}_2', \cdots, {\mathbf{k}}_{N/2}'$,
 with the constraint ${\mathbf{k}}_i'\neq {\mathbf{k}}_{j}'$
and ${\mathbf{k}}_i'\neq  {\mathbf{k}}$, where $i,j=2\cdots N/2$.
One can obtain $\langle 0_{{\mathbf{k}},s}|\rho_1({\mathbf{k}},s)
|0_{{\mathbf{k}},s}\rangle = 1-x_{\mathbf{k}}$. Hence in the basis
$(|0_{{\mathbf{k}},s}\rangle, |1_{{\mathbf{k}},s}\rangle)$,
\begin{equation}
\rho_1({\mathbf{k}},s) = diag(1-x_{\mathbf{k}},x_{\mathbf{k}}).
\end{equation}

One can also obtain that the  element of the  reduced density
matrix for one pair of modes with
 the opposite  $\mathbf{k}$ and
$s$, denoted as $\langle n_{{\mathbf{k}},s}, n_{-{\mathbf{k}},-s}
|\rho_2 ({\mathbf{k}},s;-{\mathbf{k}},-s) |
n_{{\mathbf{k}},s},n_{{-\mathbf{k}},-s}\rangle$, is
$x_{\mathbf{k}}$ when $n_{{\mathbf{k}},s}=
n_{-{\mathbf{k}},-s}=1$,  is  $1-x_{\mathbf{k}}$ when
$n_{{\mathbf{k}},s}=
 n_{-{\mathbf{k}},-s}=0$, and is $0$ otherwise. Hence in the basis
 $(|0_{{\mathbf{k}},s}0_{{\mathbf{-k}},-s}\rangle,
  |0_{{\mathbf{k}},s}1_{{\mathbf{-k}},-s}\rangle,
  |1_{{\mathbf{k}},s}0_{{\mathbf{-k}},-s}\rangle,
   |1_{{\mathbf{k}},s}1_{{\mathbf{-k}},-s}\rangle)$,
 \begin{equation}
 \rho_2 ({\mathbf{k}},s;-{\mathbf{k}},-s) =
 diag(1-x_{\mathbf{k}},0,0,x_{\mathbf{k}}). \label{re}
\end{equation}

Therefore the entanglement between the occupation-number at mode
$({\mathbf{k}},s)$  and others is
\begin{equation}
S=-x_{\mathbf{k}}\ln x_{\mathbf{k}}- (1-x_{\mathbf{k}})\ln
(1-x_{\mathbf{k}}),
\end{equation}
so is also the entanglement
between the occupation-numbers of the pair $({\mathbf{k}},s)$ and
 $(-{\mathbf{k}},-s)$ on one hand,  and the rest of the system on the
other. Note that in $|\psi(N)\rangle$, there is no entanglement
between each pair $({\mathbf{k}},s)$ and  $(-{\mathbf{k}},-s)$, as
can be simply confirmed  by the fact that (\ref{re}) is diagonal.

If $g_{\mathbf{k}}$ is $1$ for $|{\mathbf{k}}| < k_f$ and is $0$
for $|{\mathbf{k}}| > k_f$, then $x_{\mathbf{k}}$ is $1$ for
$|{\mathbf{k}}| < k_f$ and is $0$ for  $|{\mathbf{k}}| > k_f$.
Consequently for any $\mathbf{k}$, each of those  Fock-space
reduced density matrices  only has one non-vanishing element.
Therefore the entanglement $S$ reduces to zero, consistent with
the fact that under this limit, the state (\ref{nstate}) reduces
to the ground state of a free Fermi  gas~\cite{schrieffer}.

In the superconducting state, $g_{\mathbf{k}}$ differs from  that
of the free Fermi gas in the vicinity of the Fermi surface,
consequently the amount of entanglement becomes nonzero.
$g_{\mathbf{k}}$ is just the relative momentum wavefunction of
each Cooper-paired electron, and is directly related to the
superconducting energy gap $\Delta_{\mathbf{k}}$ as
$g_{\mathbf{k}}/(1+g_{\mathbf{k}}^2)
=\Delta_{\mathbf{k}}/2E_{\mathbf{k}}$, where
$E_{\mathbf{k}}=\sqrt{{\mathbf{k}}^2/2m+\Delta_{\mathbf{k}}^2}$.
As the order parameter, superconducting energy gap is a key
physical property of superconductivity.

Therefore {\em we have obtained a direct relation between
entanglement and the superconducting energy gap and thus various
physical properties of superconductivity. The entanglement
vanishes if and only if the superconductivity vanishes}.

Although superconductivity may be loosely described as Bose
condensation of Cooper pairs, it is  understood that a Cooper pair
is still different from a boson,  the strong overlap and
correlations between Cooper pairs gives rise to the gap which is
absent in the case of a Bose gas~\cite{schrieffer}. The crossover
between Bose condensation and BCS superconductivity has been an
interesting topic for a long time. Here we have found that
entanglement in $|\psi(N)\rangle$ provides a quantitative
characterizations of the correlations between Cooper pairs and
thus may be useful in studying of the crossover between Bose
condensation and superconductivity.

After this work was done, there appeared a preprint on
entanglement in BCS state involving strong
interaction~\cite{miller}.

\section*{6. Bose Systems}

Consider a system of spinless Bosons. The proper single particle
basis is the momentum state. An eigenstate of a free spinless
boson system is simply
$|n_{{\mathbf{q}}_1},n_{{\mathbf{q}}_2},\cdots\rangle
=(a_{{\mathbf{q}}_1}^{\dagger})^{n_{{\mathbf{k}}_1}}
(a_{{\mathbf{q}}_2}^{\dagger})^{n_{{\mathbf{q}}_2}}\cdots|0\rangle$.
In the ground state $(a_0^{\dagger})^N|0\rangle$, all particles
occupy the zero momentum state. This is Bose-Einstein
condensation. The system  is obviously non-entangled, in the
proper single particle basis, in all the eigenstates. Thus there
is entanglement in position basis, in consistent with a related
work~\cite{simon}.

For a weakly interacting spinless boson gas,  entanglement between
occupation-numbers of different momentum states is still absent
under Hatree-Gross-Pitaevskii approximation. In this approach, an
energy eigenstate is approximated as a product of single particle
states, with symmetrization, hence there is no occupation-number
entanglement. The ground state is a product of a same single
particle state. The weak interaction only affects the single
particle state. Nevertheless, there may be entanglement when there
is spin degree of freedom or in other complex
situations~\cite{shi3,milburn}. These features are like those of
the Hatree-Fock approach of a Fermi gas.

The next level of treatment is Bogoliubov theory~\cite{bogo},
nonzero entanglement appears, even in the ground state. It is
first hinted by the Bogoliubov transformation in the original,
particle non-conserving, formulation,  which defines a new
annihilation operator which is a superposition of a annihilation
operator $a_{\mathbf{q}}$ and the creation operator for the
opposite momentum $a_{-\mathbf{q}}^{\dagger}$,
namely, $b_{\mathbf{q}}=
u_{\mathbf{q}}a_{\mathbf{q}}+v_{\mathbf{q}}a_{-\mathbf{q}}^{\dagger}$.
This transformation diagonalizes the second quantized Hamiltonian,
hence in terms of the newly defined quasiparticles, there is no
entanglement, signalling that there exists entanglement in terms
of the original particles.  Similar to BCS superconductivity, in
the particle non-conserving theory, entanglement only exists
between the each pair of modes $\mathbf{q}$ and $-\mathbf{q}$
($\mathbf{q} > 0$). The ground state is~\cite{huang,laughlin2}
\begin{equation}
\begin{array}{cl}
|\Psi_0\rangle \propto &
\sum_{n_{{\mathbf{q}}_1}}\sum_{n_{{\mathbf{q}}_2}}\cdots
[(-v_{{\mathbf{q}}_1}/u_{{\mathbf{q}}_1})^{n_{\mathbf{q}_1}}
(-v_{{\mathbf{q}}_2}/u_{{\mathbf{q}}_2})^{n_{{\mathbf{q}}_2}}\cdots]
\\
&|n_0;n_{\mathbf{q}_1},n_{\mathbf{q}_1};n_{{\mathbf{q}}_2},n_{{\mathbf{q}}_2};
\cdots\rangle,
\end{array} \label{bogo}
\end{equation}
in which there are $n_0$ particles with zero momentum while
$n_{\mathbf{q}}$ pairs of particles with $\mathbf{q}$ and
$-\mathbf{q}$. Therefore, the entanglement between occupation
numbers at $\mathbf{q}$ and $-\mathbf{q}$ is $S=-\sum_i x_i\ln
x_i$, where $x_i=y_i/\sum_i y_i$, $n=0,1,\cdots,\infty$. where
$y_i=|v_{\mathbf{q}}/u_{\mathbf{q}}|^{2i}$.  The condensate mode
is indeed disentangled from the rest of the system, in consistent
with our result obtained from ODLRO.

Vedral studied entanglement in a Bose condensate, using a state
similar to Eq.~(\ref{bogo}), and calculated a different quantity
defined there to measure the amount of
entanglement~\cite{vedral1}.

We now focus on the particle number conserving version of the
Bogoliubov theory, which gives the ground state as~\cite{leggett}
\begin{equation}
|\Psi(N)\rangle \propto
(a_0^{\dagger}a_0^{\dagger}-\sum_{|\mathbf{q}|\neq 0}
c_{\mathbf{q}}a_{\mathbf{q}}^{\dagger}a_{-\mathbf{q}}^{\dagger})^{N/2}
|0\rangle,
\end{equation}
where $c_{\mathbf{q}}$, with $|c_{\mathbf{q}}|< 1$ is determined
by the Hamiltonian and is the effect of the weak interaction. It
can be found that
\begin{equation}
\begin{array}{l}
|\Psi(N)\rangle \propto  \sum\limits_{n_0,n_1\cdots n_{\infty}}
p(N/2;n_0\cdots n_{\infty}) (-c_{\mathbf{q}_1})^{n_1}\cdots\\
\times (-c_{\mathbf{q}_{\infty}})^{n_{\infty}}
|2n_0\rangle_0|n_1\rangle_{\mathbf{q}_1}|n_1\rangle_{-\mathbf{q}_1}
\cdots|n_{\infty}\rangle_{\mathbf{q}_{\infty}}|n_{\infty}\rangle_{-\mathbf{q}_{\infty}}
\end{array}
\end{equation}
where $n_0+n_1+\cdots+n_{\infty}=N/2$,
$p(N/2;n_0\cdots n_{\infty})=\frac{(N/2)!}{n_0!n_1!\cdots
n_{\infty}!}$ is the number of partitions of  $N/2$ objects into
different boxes, with $n_0$ objects in the box labelled $0$, $n_1$
in the box labelled $1$, and so on. Then one obtains the
Fock-space reduced density matrices of different momentum states.
The nonvanishing elements of $\rho_1(0)$ are
\begin{equation}
\begin{array}{l}
x_{2n_0} (0) \equiv \langle 2n_0 |\rho_1(0)|2n_0 \rangle  \\
= A \sum\limits_{n_1\cdots n_{\infty}}
p^2(N/2-n_0;n_1\cdots n_{\infty})|c_{\mathbf{q}_1}|^{2n_1}\cdots
|c_{\mathbf{q}_{\infty}}|^{2n_{\infty}},
\end{array}
\end{equation}
where $n_1+\cdots + n_{\infty}=N/2-n_0$, the normalization factor
$A= [\sum_{n_0,\cdots, n_{\infty}}
p^2(N/2;n_0,\cdots,n_{\infty})|c_{\mathbf{q}_1}|^{2n_1}\cdots
|c_{\mathbf{q}_{\infty}}|^{2n_{\infty}}]^{-1}$, with
$n_0+n_1+\cdots + n_{\infty}=N/2$. It can be seen that
\begin{equation}
x_{2n_0}(0) \approx \frac{|\sum\limits_{|\mathbf{q}|\neq 0}
c_{\mathbf{q}}|^{N-2n_0}}{\sum\limits_{n_0=0}^{N/2}
|\sum\limits_{|\mathbf{q}|\neq
0} c_{\mathbf{q}}|^{N-2n_0}}, \label{x0}
\end{equation}
under the assumption that $\sum c_{\mathbf{q}_1}^*c_{\mathbf{q}_2}
\approx 0$, where the summation is over all nonzero $\mathbf{q}_1
\neq \mathbf{q}_2$.

The entanglement between the zero momentum state and the rest of
the system is
\begin{equation}
S(0)= -\sum_{n_0=0}^{N/2} x_{2n_0} (0) \ln x_{2n_0} (0).
\end{equation}

For a momentum $\mathbf{q}_1 \neq 0$, the nonvanishing elements of
$\rho_1(\mathbf{q}_1)$ are
\begin{equation}
\begin{array}{l}
x_{n_1} (\mathbf{q}_1) \equiv \langle n_1
|\rho_1(\mathbf{q}_1)|n_1 \rangle \\
= A |c_{\mathbf{q}_1}|^{2n_1} \sum\limits_{n_0,n_2\cdots
n_{\infty}}
p^2(N/2;n_0\cdots n_{\infty})|c_{\mathbf{q}_1}|^{2n_2}\cdots
|c_{\mathbf{q}_{\infty}}|^{2n_{\infty}},
\end{array}
\end{equation}
where the summation is subject to $n_0+n_2+\cdots +
n_{\infty}=N/2-n_1$. It can be seen that
\begin{eqnarray}
x_{n_1}  (\mathbf{q}_1) \approx
\frac{|c_{\mathbf{q}_1}|^{2n_1}\sum\limits_{n_0=0}^{N/2-n_1}
|\sum\limits_{|\mathbf{q}|\neq 0,\mathbf{q}_1}
c_{\mathbf{q}}|^{N-2n_0-2n_1}}{\sum\limits_{n_1=0}^{N/2}
|c_{\mathbf{q}_1}|^{2n_1}\sum\limits_{n_0=0}^{N/2-n_1}
|\sum\limits_{|\mathbf{q}|\neq 0,\mathbf{q}_1}
c_{\mathbf{q}}|^{N-2n_0-2n_1}} \nonumber \\
 = \frac{ |c_{\mathbf{q}_1}|^{2n_1}-
(\frac{|c_{\mathbf{q}_1}|}{|\sum\limits_{|\mathbf{q}|\neq
0,\mathbf{q}_1} c_{\mathbf{q}}|})^{2n_1}
|\sum\limits_{|\mathbf{q}|\neq
0,\mathbf{q}_1}c_{\mathbf{q}}|^{N+1}}
{\sum\limits_{n_1=0}^{N/2}[|c_{\mathbf{q}_1}|^{2n_1}-
(\frac{|c_{\mathbf{q}_1}|}{|\sum\limits_{|\mathbf{q}|\neq
0,\mathbf{q}_1}
c_{\mathbf{q}}|})^{2n_1}|\sum\limits_{|\mathbf{q}|\neq
0,\mathbf{q}_1} c_{\mathbf{q}}|^{N+1}]} \label{x1}
\end{eqnarray}

The entanglement between a non-zero momentum state $\mathbf{q}_1$
and the rest of the system is
\begin{equation}
S(\mathbf{q}_1)= -\sum_{n_1=0}^{N/2} x_{n_1} (\mathbf{q}_1) \ln
x_{n_1} (\mathbf{q}_1).
\end{equation}

Obviously, the entanglement between $-\mathbf{q}_1$ and the rest
of the system, as well as the entanglement between the pair
$\mathbf{q}_1$ plus $-\mathbf{q}_1$ and the rest of the system,
are both also $S(\mathbf{q}_1)$.  It can also be seen that there
is no entanglement between  $\mathbf{q}_1$ and $-\mathbf{q}_1$. On
this aspect, there is a similarity with BCS superconductivity.

Consider the identity $\sum_{n_{\mathbf{q}}} n_{\mathbf{q}}
\langle n_{\mathbf{q}} |\rho_1(\mathbf{q})|n_{\mathbf{q}} \rangle
= {{\sum'}_{\{n_i\}}} n_{\mathbf{q}} \langle n_0 \cdots
n_{\infty}|\rho|n_0 \cdots n_{\infty}\rangle
 = \langle \hat{n}_{\mathbf{q}} \rangle$,
for  $n_{\mathbf{q}}=0,1, \cdots$, and for different
$\mathbf{q}$'s, where ${{\sum'}_{\{n_i\}}}$ represents summations
over $n_0,\cdots,n_{\infty}$, except $n_{\mathbf{q}}$. For the
Bogoliubov ground state $|\Psi(N)\rangle$, it is known that
$\langle \hat{n}_0 \rangle$ is close to $N$, while $\langle
\hat{n}_{\mathbf{q}_1} \rangle \ll N$ for $\mathbf{q}_1 \neq 0$.
It is thus constrained that only a small number (compared with
$N$) of the Fock space matrix elements $\langle n_{\mathbf{q}}
|\rho_1(\mathbf{q})| n_{\mathbf{q}} \rangle$ is considerable for
mode $\mathbf{q}$. Thus the entanglement is small. But it is not
zero, as in the Hatree approximation.

Furthermore, (\ref{x0}) indicates that $x_{2n_0}(0)$ exponentially
decays with the $n_0$, with the rate
$1/2\ln|\sum_{|\mathbf{q}|\neq 0} c_{\mathbf{q}}|$. Hence indeed
very small number of matrix elements $\langle n_{0} |\rho_1(0)|
n_{0} \rangle$ is considerable, and thus $S(0)$ is small. On the
other hand, in (\ref{x1}), the change of $x_{n_1}(\mathbf{q}_1)$
with $n_1$ is slower since it involves the counteracting of two
exponentially increasing terms. Consequently $S(\mathbf{q}_1)
> S(0)$.

The small but nonzero entanglement is a characterization of the
two-particle correlation caused by the weak interaction, which is
the essence of Bogoliubov theory~\cite{leggett}. This can be seen
from Eqs.~(\ref{x0}) and (\ref{x1}), which indicates that the
entanglement is only dependent on the function $c_{\mathbf{q}}$,
which is determined by the weak interaction.

The result in last and this section is the entanglement in terms
of the original particles. As consistent with the fact that
entanglement depends on which single particle is used in
representing the many-particle system, it can be shown that in the
set of eigenstates of the one-particle reduced density matrix (in
the case of Bose condensation) or two-particle reduced density
matrix (in the case of superconductivity),  the condensate mode is
disentangled with the rest of the system~\cite{dis}.

\section*{7. Summary and remarks}

It is known that quantum correlation in a system of identical
particles can be characterized in terms of entanglement between
occupation-numbers of different single particle basis states, and
thus depends on which single particle basis  is chosen.
Consequently,  in general, occupation-number entanglement may be
generated, or exists in the energy eigenstates,  even in absence
of interaction of  particles. Indeed, it is caused by coupling
between different single particle basis states (mode-mode
coupling) in the many-particle second-quantized Hamiltonian, which
may exist even in the one-body term in the Hamiltonian.

However, our purpose in this paper is to use entanglement as a
characterization of effects of interaction. For this purpose, we
choose the set of eigenstates of the single particle Hamiltonian
as the single particle basis on which the entanglement is defined.
For convenience, we call it proper single particle basis. In this
single particle basis, if there is no interaction, the
second-quantized Hamiltonian is diagonal in different single
particle basis states, and thus the many-particle eigenstates can
always be chosen to be non-entangled.

Using the so-called proper single particle basis state, we
examined entanglement in eigenstates, especially the ground
states, of some important many-particle Hamiltonians. These
examples demonstrate that entanglement in the proper single
particle basis can indeed characterize the effect of interaction,
vanishing as the interaction vanishes. Moreover, the amount of the
entanglement calculated is directly related to the macroscopic
physical properties. In other words, it is demonstrated that the
microscopic entanglement is manifested in the macroscopic physical
properties. It appears that entanglement in the proper single
particle basis is useful especially for studying the strongly
correlated systems, in which interactions are important.

For an interacting Fermi gas, electron-hole entanglement
inevitably appears in some excited eigenstates, as described in
the Hatree-Fock approach.  Electron-hole entanglement can be
generated by electron-light interaction, which is not included in
the single particle Hamiltonian which defines the proper single
particle basis.

When the ground state of a Fermi liquid is expressed in terms of
the Landau quasiparticles, i.e. electrons dressed by the
interaction, it becomes non-entangled.

We found the nice result that the entanglement in a quantum Hall
state is just the entropy of the probability distribution $f$ and
$1-f$, where $f$ is the proper fractional part of the filling
factor of a Landau level. Hence entanglement here can be extremely
precisely measured, with topological stability. This gives a
support to the well-known proposal of using anyons for
fault-tolerant quantum computing.

We also made a detailed calculation of entanglement in BCS ground
state. Both the particle-number non-conserved and the particle
number conserved  states are considered. In each case,  the amount
of entanglement is a function of the relative momentum
wavefunction $g_{\mathbf{k}}$ of every two Cooper-paired
particles, and thus directly related to the superconducting energy
gap. The entanglement vanishes if and only if  the
superconductivity vanishes.

Finally, we turned to Bose systems. For a spinless system, the
entanglement is absent in the eigenstates in  the
Hatree-Gross-Pitaevskii approximation.  However,  though small, it
is non-vanishing in  the Bogoliubov theory, using which  we
calculate the entanglement in the ground state, where there is a
kind of pairing between opposite momenta. Entanglement in the
proper single particle basis provides a characterization of the
two-particle correlation due to interaction, which is the essence
of Bogoliubov theory.

Many-body entangled states as those in condensed matter physics
may be useful for quantum information processing. One may
adiabatically control a time-dependent many-body  ground state
which encodes the quantum information. If there is a finite energy
gap between the ground state and the excited states, as existing
in many condensed matter systems, such a quantum information
processing should naturally possess some robustness against
environmental perturbation.

But more caution is needed in using entanglement in a condensed
matter system to demonstrate Bell theorem and such. A reason is
that in condensed matter physics, many Hamiltonians, usually
instantaneous, are effective ones on a certain time scale, with
many degrees of freedom renormalized. The formal entangled state
and the instantaneous correlations may be meaningful only on a
certain coarse-grained time scale.

For identical particles, there is intrinsic built-in
non-separability because of the pre-condition that the spatial
wavefunctions overlap, for example, spin magnetism based on
exchange interaction originates in antisymmetrizing the spin-orbit
states of the electrons interacting with ``instantaneous'' Coulomb
interaction which is always there. Deeper understanding is still
needed on the occupation-number entanglement.

\section*{Acknowledgments}

I thank Professors Tony Leggett, Peter Littlewood, John Preskill,
William Wootters, Yong-Shi Wu and Chen Ning Yang for useful
discussions.


\begin{thebibliography}{99}

\bibitem{einstein} E. Einstein, B. Podolsky and N. Rosen, Phys. Rev.
{\bf 47} 777 (1935).  E. Schr\"{o}dinger, Proc. Camb. Phi. Soc.
{\bf 31}, 555 (1935). D. Bohm, {\em Quantum Theory}
(Prentice-Hall, Englewood cliffs, 1951).

\bibitem{bell} J. S. Bell, Physics {\bf 1}, 195 (1964).

\bibitem{mermin} N. D. Mermin, Rev. Mod. Phys. {\bf 65}, 803  (1993).
A. Peres, {\em Quantum Theory: Concepts and Methods} (Kluwer Academic,
Dordrecht, 1993).

\bibitem{bennett} C. H. Bennett, H. J. Bernstein, S. Popescu and
B. Schumacher, Phys. Rev. A {\bf 53}, 2046 (1996).

\bibitem{yang} C. N. Yang, Rev. Mod. Phys. {\bf 34}, 694 (1962).

\bibitem{leg} A. J. Leggett, Prog. Theor. Phys. Suppl. {\bf 69}, 80 (1980).

\bibitem{preskill} J.~Preskill, J. Mod. Opt. {\bf 47}, 127 (2000).

\bibitem{hei1} K. M. O'Connor and W. K. Wootters,
Phys. Rev. A {\bf 63}, 052302 (2001).

\bibitem{thermal}
M. C. Arnesen, S. Bose and V. Vedral, Phys. Rev. Lett. {\bf 87},
017901 (2001); M. A. Nielsen,  quant-ph/0011036.

\bibitem{osterloh} A. Osterloh {\it et
al.}, Nature, {\bf 416}, 608 (2002).

\bibitem{osborn} T. J. Osborne and M. A. Nielsen, Phys. Rev. A {\bf 66}, 032110
(2002).

\bibitem{vidal} G. Vidal, J. I. Latorre, E. Rico and A. Kitaev, quant-ph/0211074;
J. I. Latorre, E. Rico and G. Vidal, quant-ph/0304098.

\bibitem{korepin1} B. Q. Qin and V. Korepin, quant-ph/0304108.
L. Amico and A. Osterloh, cond-mat/0306285. B. Q. Qin and V.
Korepin, quant-ph/0309188.

\bibitem{korepin2} V. Korepin, quant-ph/0311056.

\bibitem{abrams} D. S. Abrams and S. Lloyd, Phys. Rev. Lett. {\bf 79}, 2586
(1997).

\bibitem{lloyd1} S. Lloyd, in {\em Unconventional Models of Computation},
 C.S. Calude, J. Casti, M.J. Dinneen, eds. (Springer, Singapore,
 1998), quant-ph/0003151.

\bibitem{lloyd2} S. Lloyd, Quantum Information Processing {\bf 1},
13 (2002).

\bibitem{bravyi} S. Bravyi and A. Kitaev, quant-ph/0003137.

\bibitem{mn} E. Knill, R. Laflamme and G. J. Milburn, Nature (London) {\bf 409}, 46
(2001).

\bibitem{terhal} B. M. Terhal and D. P. DiVincenzo, Phys. Rev. A {\bf 65}, 032325 (2002).

\bibitem{zanardi1} P. Zanardi, Phys. Rev. A. {\bf 65}, 042101
(2002). P. Zanardi, X. Wang, J. Phys. A, 35 (2002) 7947.

\bibitem{zanardi0} P. Zanardi,
Phys. Rev. A 67, 054301 (2003). P. Giorda and P. Zanardi,
quant-ph/0304151.  P. Zanardi and S. Lloyd, quant-ph/0305013. P.
Giorda and  P. Zanardi, quant-ph/0311058. P. Zanardi, D. Lidar and
S.Lloyd, Phys. Rev. Lett. {\bf 92}, 060402 (2004).

\bibitem{shi4} Y. Shi, Phy. Rev. A {\bf 67}, 024301 (2003).

\bibitem{shi0} An earlier version of the present work was
described in a preprint [Y. Shi, quant-ph/0204058].

\bibitem{dis} Y.~Shi, Phys. Lett. A {\bf 309}, 254 (2003).

\bibitem{enk}  S. J. van Enk, Phys. Rev. A {\bf 67}, 022303 (2003).

\bibitem{vedral1} V. Vedral, Central Eur. J. Phys. {\bf 1}, 289 (2003).

\bibitem{ghirardi} G. Ghirardi, L. Marinatto and  T. Weber,
J. Stat. Phys. {\bf 108}, 49 (2002)

\bibitem{knill} H. Barnum, E. Knill, G. Ortiz and L. Viola, quant-ph/0207149;
H. Barnum,  E. Knill, G. Ortiz, R. Somma and L. Viola,
quant-ph/0305023.

\bibitem{schliemann} J. Schliemann {\it et al.}, Phys. Rev. A, {\bf 64},
022303 (2001).

\bibitem{zeh} A particle in a classical sense can emerge as a consequence of decoherence
[cf. H. D. Zeh, Phys. Lett. A {\bf 309}, 329 (2003)].

\bibitem{pais} A. Pais, {\em Subtle is the Lord} (Oxford
University, Oxford, 1982).

\bibitem{farhi} E. Farhi {\it et al.},  Science {\bf 292}, 472 (2001).
A. M. Childs, E. Farhi and J. Preskill, Phys. Rev. A {\bf 65},
012322 (2002).

\bibitem{zanardi2}
P. Zanardi and M. Rasetti, Phys. Lett. A {\bf 264}, 94 (1999).

\bibitem{sw} Y. Shi and Y. S. Wu, Phys. Rev. A {\bf 69}, 024301 (2004).

\bibitem{solid} For example, G. Grosso and G. P. Parravichi, {\em
Solid State Physics} (Academic Press, San Diego, 2000).

\bibitem{chen} G. Chen {\it et al.}, Science {\bf 289}, 1906
(2000).

\bibitem{sl} Y. Shi, Phys. Rev. A {\bf 69}, 032318 (2004).

\bibitem{landau} L. D. Landau, Soviet Phys. JETP {\bf 3}, 920
(1957).

\bibitem{luttinger} J. M. Luttinger, J. Math. Phys. {\bf 4}, 1154 (1963).

\bibitem{qhe} For example, R. E. Prange and S. M. Girvin,
{\em  The Quantum Hall Effect} (Springer, Berlin, 1990);  S. Das
Sarma and A. Pinczuk, {\em Perspectives in Quantum Hall Effects}
(Wiley, 1997); Z. F. Ezawa, {\em Quantum Hall Effect} (World
Scientific, Singapore, 2000).

\bibitem{laughlin} R. B. Laughlin, Phys. Rev. Lett. {\bf 50}, 1395 (1983).

\bibitem{girvin} S. M. Girvin and A. H. MacDonald, Phys. Rev. Lett.
{\bf 58}, 1252 (1987).

\bibitem{kitaev} A. Yu. Kitaev, Annals Phys. {\bf 303}, 2 (2003).

\bibitem{bcs} J. Bardeen, L. N. Cooper and J. R. Schrieffer, Phys.
Rev. {\bf 108}, 1175 (1957).

\bibitem{schrieffer} J. R. Schrieffer, {\em Theory of
Superconductivity} (Perseus Books, Reading, 1964).

\bibitem{simon} C. Simon, Phys. Rev. A {\bf 66}, 052323 (2002).

\bibitem{miller} D. Miller and Abdel-Nasser M. Tawfik,
hep-ph/0312368.

\bibitem{shi3} Y.~Shi, Int. J. Mod. Phys. B {\bf 15}, 22 (2001).

\bibitem{milburn}  A. P. Hines, R. H. McKenzie and G. J. Milburn,
Phys. Rev. A {\bf 67}, 013609 (2003).

\bibitem{bogo} N. N. Bogoliubov, J. Phys. (USSA) {\bf 11}, 23
(1947).

\bibitem{huang} K. Huang, {\em Statistical Mechanics} (Wiley, New York,
1987)

\bibitem{laughlin2}  Entanglement in Bogoliubov state is
qualitatively noted in A. Granik and G. Chapline, quant-ph/0302013
and R. B. Laughlin, gr-qc/0302028.

\bibitem{leggett} A. J. Leggett,  Rev. Mod. Phys.~{\bf 73}, 307 (2001).


\end{thebibliography}
\end{document}